\documentclass[sigconf, nonacm]{acmart}

\AtBeginDocument{%
  }

\newcommand{\thiswork}{\textsc{BIP! Scholar}} 

\usepackage{multirow}
\usepackage[normalem]{ulem}

\begin{document}

\title{\thiswork: A Service to Facilitate Fair Researcher Assessment}

\author{Thanasis Vergoulis}
\orcid{0000-0003-0555-4128}
\affiliation{%
  \institution{IMSI, ATHENA RC}
  \city{Athens}
  \country{Greece}
}
\email{vergoulis@athenarc.gr}

\author{Serafeim Chatzopoulos}
\orcid{0000-0003-1714-5225}
\affiliation{%
  \institution{IMSI, ATHENA RC}
  \city{Athens}
  \country{Greece}
}
\email{schatz@athenarc.gr}

\author{Kleanthis Vichos}
\orcid{0000-0002-8955-9489}
\affiliation{%
  \institution{IMSI, ATHENA RC}
  \city{Athens}
  \country{Greece}
}
\email{kvichos@athenarc.gr}

\author{Ilias Kanellos}
\orcid{0000-0003-2146-3795}
\affiliation{%
  \institution{IMSI, ATHENA RC}
  \city{Athens}
  \country{Greece}
}
\email{ilias.kanellos@athenarc.gr}

\author{Andrea Mannocci}
\orcid{0000-0002-5193-7851}
\affiliation{%
  \institution{ISTI, CNR}
  \city{Pisa}
  \country{Italy}
}
\email{andrea.mannocci@isti.cnr.it}

\author{Natalia Manola}
\orcid{0000-0002-3477-3082}
\affiliation{%
  \institution{OpenAIRE AMKE}
  \city{Athens}
  \country{Greece}
}
\email{natalia.manola@openaire.eu}

\author{Paolo Manghi}
\orcid{0000-0001-7291-3210}
\affiliation{%
  \institution{ISTI, CNR}
  \city{Pisa}
  \country{Italy}
}
\additionalaffiliation{
  \institution{OpenAIRE AMKE, Athens, Greece}
}
\email{paolo.manghi@isti.cnr.it}

\renewcommand{\shortauthors}{Vergoulis and Chatzopoulos, et al.}

\begin{abstract}
In recent years, assessing the performance of researchers has become a burden due to the extensive volume of the existing research output. As a result, evaluators often end up relying heavily on a selection of performance indicators like the h-index. However, over-reliance on such indicators may result in reinforcing dubious research practices, while overlooking important aspects of a researcher's career, such as their exact role in the production of particular research works or their contribution to other important types of academic or research activities (e.g., production of datasets, peer reviewing). In response, a number of initiatives that attempt to provide guidelines towards fairer research assessment frameworks have been established. In this work, we present~\thiswork, a Web-based service that offers researchers the opportunity to set up profiles that summarise their research careers taking into consideration well-established guidelines for fair research assessment, facilitating the work of evaluators who want to be more compliant with the respective practices.
\end{abstract}


\maketitle

\section{Introduction}

Advancements in the career of a researcher (recruitment, promotion, etc.) and professional achievements (for instance, awarding of prizes or funds) essentially rely on some type of performance assessment. Ideally, this involves scrutinising the curriculum vitae of the researcher of interest (and of their competitors, if this applies). However, nowadays, this process can be extremely tedious due to the extensive productivity of modern researchers, driven by the shifting nature of the scientific paradigm~\cite{gray2009} and by the intense competition among them, a notorious trend widely known as ``publish or perish''~\cite{pop}. Consequently, in an attempt to reduce this burden, evaluators turn to the use of evaluation ``shortcuts''~\cite{ten-ways}, i.e., indicators such as the number of papers, the 
impact factor of the journals where they publish them, their citations, etc. 

However, relying heavily on such indicators can amplify important problems (e.g., the Matthew effect~\cite{matthew}) or even foster bad research practices ~\cite{ten-ways}. In addition, most platforms that provide relevant data (e.g., Google Scholar, ResearchGate) offer a limited variety of indicators (usually only the number of publications and a couple of citation-based metrics, like
the h-index) without providing adequate explanations for their interpretation and proper use. To make matters worse, these platforms almost solely focus on publications, failing to acknowledge contributions to other types of research works (e.g., datasets, software) that consume significant effort by the researchers and are important as well. 
The same holds for additional research activities that are essential for safeguarding and advancing science, like peer-review and data/code sharing~\cite{advancing}. Finally, the aforementioned indicators are completely oblivious of the different types of roles that researchers may have in a research project~\cite{hidden-roles}. Due to these issues, it is not surprising that promotion and tenure processes in various institutions do not reward open and responsible research practices~\cite{pontika2022indicators}, while they 
additionally affect the way in which researchers conceive and conduct research~\cite{advancing}.
\begin{figure*}[t]
    \centering
    \fbox{\includegraphics[width=0.8\textwidth]{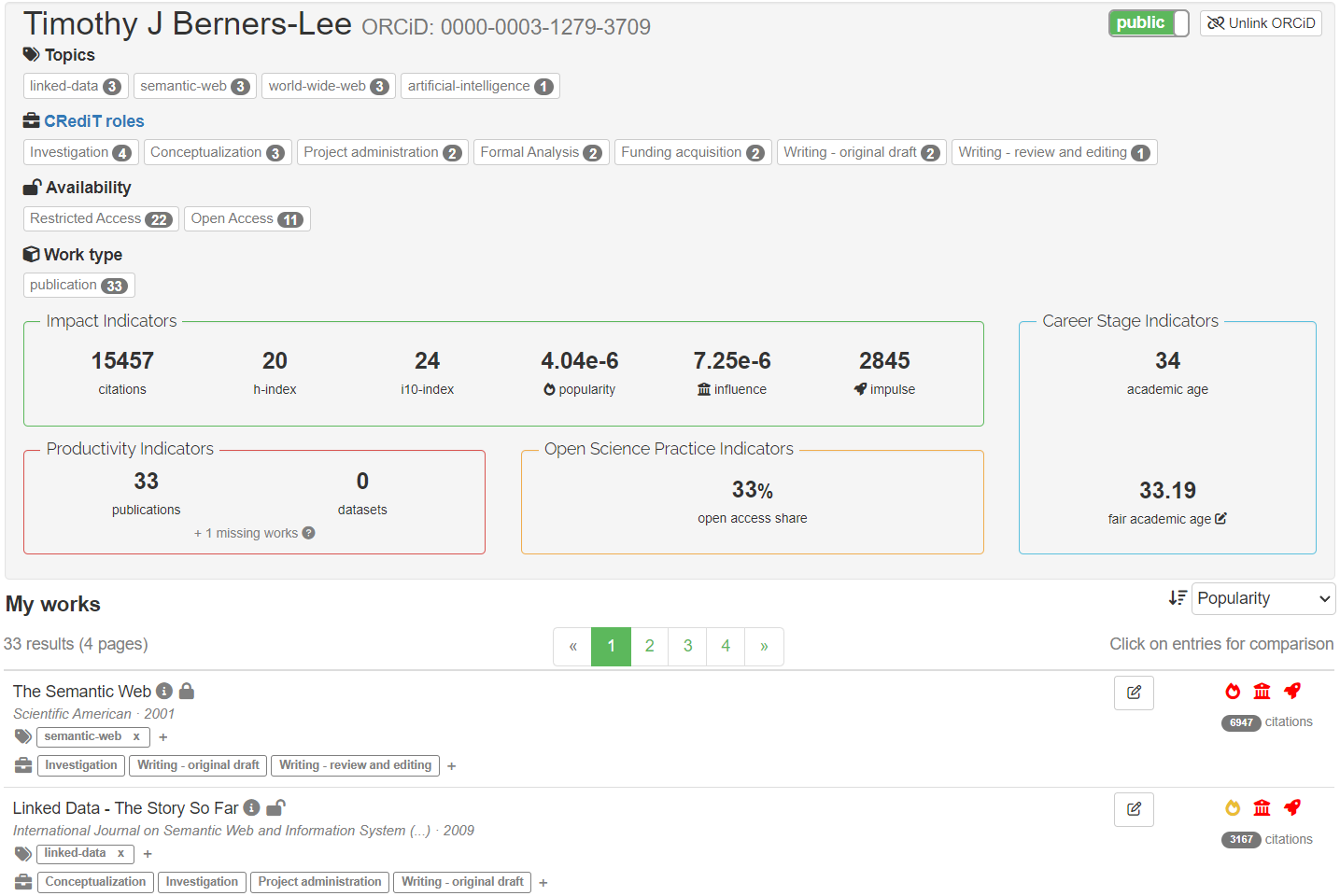}}
  \caption{An indicative profile of a well-known scientist.}
  \label{fig:ui}
\end{figure*}
As a reaction to the above, various initiatives (e.g., DORA\footnote{DORA: \url{https://sfdora.org}}, the Leiden Manifesto~\cite{leiden}, the Hong Kong Principles~\cite{hk}), which attempt to motivate a reform in the current research assessment frameworks 
employed by research organisations and funding institutions, have been developed. These initiatives aim to provide useful guidelines towards a fairer evaluation of researchers. For instance, DORA advocates, among others, in favour of abandoning the use of journal-based metrics to assess researchers, using 
instead a variety of article-level indicators, and taking into consideration the value of research datasets and software. 
In addition, other initiatives highlight the importance of acknowledging the compliance of researchers with open science practices (e.g., OS-CAM~\cite{oscam}). Finally, many initiatives as such acknowledge that, today, research teams are heterogeneous, consisting of and relying on people having diverse specialisations and roles (e.g., some being better in conducting experiments, others in writing papers etc). However, currently, researchers rarely have the opportunity to indicate such roles (not all editors support this) and, consequently, the current assessment system does not take this perspective into consideration. 

Inspired by these initiatives, we have developed \thiswork\footnote{\thiswork: \url{https://bip.imis.athena-innovation.gr/scholar/profile}}, a Web-based service that aims to facilitate the work of evaluators who want to adopt various fair assessment practices for researchers (details in Section~\ref{sec:profile}). \thiswork~leverages ORCID profiles and the CRediT taxonomy\footnote{CRediT – Contributor Roles Taxonomy: \url{https://casrai.org/credit}} to allow  researchers to better represent their academic profiles, 
capturing the different types of research works and declaring the exact role they covered in each of them.

\section{\thiswork's Researcher Profile}
\label{sec:profile}

In this section, we outline \thiswork's researcher profile functionalities (Section~\ref{sec:profile-overview}) and we elaborate on their compliance with various fair researcher assessment practices (Section~\ref{sec:practices}). 

\subsection{Overview}
\label{sec:profile-overview}

\begin{table*}[t]
  \scriptsize
  \caption{Researcher-level indicators.}
  \label{tbl:indicators}
  \begin{tabular}{ccl}
    \toprule
    Aspect & Indicator & Description\\
    \midrule
    \multirow{6}{*}{Impact} & Citations & The total number of citations received by all articles of the researcher of interest.\\
    \multirow{6}{*}{} & h-index~\cite{hirsch2005index} & It is an estimation of the importance, significance, and broad impact of a researcher's cumulative research contributions.\\
    \multirow{6}{*}{} & i10-index & This measure, introduced by Google Scholar, gauges the productivity of researchers by counting
    their number of papers with 10 or more citations.\\
    \multirow{6}{*}{} & Popularity & The sum of the popularity (current impact/attention) scores of all articles of a researcher of interest (popularity is based on AttRank~\cite{kanellos2020ranking}).\\
    \multirow{6}{*}{} & Influence & The sum of the influence (overall impact) scores of all articles of a researcher of interest. Influence scores are calculated using PageRank~\cite{page1999pagerank}.\\
    \multirow{6}{*}{} & Impulse & The sum of the impulse scores, (calculated according to the citations received in the immediate years after publication) of all articles of a researcher of interest.\\
    \midrule
    \multirow{2}{*}{Productivity} & Publications & The total number of a researcher's articles.\\
    \multirow{2}{*}{} & Datasets & The total number of a researcher's datasets.\\
    \midrule
    Open science practice & Open access share & The share (proportion) of open access articles of the researcher of interest.\\
    \midrule
    \multirow{2}{*}{Career Stage} & Academic age & It reflects the time that a scientist has spent conducting active research.\\
    \multirow{2}{*}{} & Fair academic age & A variant of the academic age indicator that takes into consideration a researcher's inactive periods.\\
  \bottomrule
\end{tabular}
\end{table*}

\thiswork~is a service that allows researchers to set up an academic CV (i.e., a profile page), which represents their research activities in detail and highlights different aspects of their research career. Registered researchers can create their own profile pages in the service by authorising it to gather their public ORCID records. Then, \thiswork~maps these records to those in its internal database, which contains metadata and a set of calculated indicators for more than $140$M research works (details in Section~\ref{sec:data}). 
It should be noted that \thiswork~considers not only publications, but also other types of research works (currently, only datasets, but in the future additional types like peer reviews and research software will be supported). 
These data are used to assembly the respective profile pages (an indicative example is presented in Figure~\ref{fig:ui}). It should be noted that each researcher can select either to keep the profile private (e.g., to use it only for self-monitoring reasons) or to make it public (allowing access to third parties).

The researcher's profile page is organised into two basic blocks: (a)~a \emph{career summary} of the researcher (the grey box on top, in Figure~\ref{fig:ui}), and (b)~the complete \emph{track record} of all of their works (a paginated list at the bottom). Each entry in the track record displays useful metadata for the respective work, which have been collected or calculated by \thiswork~(e.g., its title, venue/repository, availability, various impact indicators). In addition, each entry contains a set of user-provided metadata: the roles of the researcher in the work (based on the classes defined by the CRediT taxonomy) and the topics which are related to it. 
Researchers are responsible for editing the roles and topics of each entry in their own profile, but they do not have permissions for the entries of other researchers. 

The career summary block, apart from basic information about the respective researcher (e.g., their name and ORCID identifier), contains a set of interesting ``faceted''~\cite{facets} classifications of their works (by topic, contribution role, availability, and type of research work), and a variety of researcher-level performance indicators. The facets are clickable and act as filters for the works in the track record. Selecting a particular combination of facets also updates the values of the indicators, making possible, for instance, to inspect the performance of the researcher in different fields or roles (each time the indicators are calculated using only the visible elements of the track record).

The researcher-level indicators offered have been selected so  as to capture a variety of aspects of a researcher's career. More specifically, \thiswork~offers different indicators that reflect the productivity, the impact, the level of practice of Open Science principles, and the career stage of researchers. To avoid misconceptions for and misuses of these indicators, the name of each indicator is a hyperlink that redirects the user to a page providing detailed explanations of the its intuition, calculation process, and known limitations, along with relevant references (if applicable). Table~\ref{tbl:indicators} summarises the implemented indicators, the performance aspect they reflect and brief explanations about their intuition. 

\subsection{Compliance with fair assessment practices}
\label{sec:practices}

The outline of the researcher profile page has been designed taking into consideration the guidelines from various research assessment initiatives to facilitate fair assessment. For instance, from the four guidelines of DORA\footnote{DORA declaration: \url{https://sfdora.org/read/}} (which is the most widely known initiative for fair research assessment) for organisations that supply metrics: 

\begin{itemize}
    \item ``Be open and transparent by providing data and methods used to calculate all metrics.''
    \item ``Provide the data under a licence that allows unrestricted reuse, and provide computational access to data, where possible.''
    \item ``Be clear that inappropriate manipulation of metrics will not be tolerated; be explicit about what constitutes inappropriate manipulation and what measures will be taken to combat this.''
    \item ``Account for the variation in article types (e.g., reviews versus research articles), and in different subject areas when metrics are used, aggregated, or compared.''
\end{itemize}

\thiswork~complies with all of them (at least to an extent). More specifically, regarding the first and third guideline, the ``Indicators'' page of the user interface provides details on the calculation of each indicator (i.e., the process followed and data used) as well as its limitations and potential misuses. 
As regards the second one, all indicator values can be accessed through our public API\footnote{BIP! API documentation: \url{https://bip-api.imsi.athenarc.gr/documentation}} under the Creative Commons Attribution 4.0 International license. 
To this end, we facilitate programmatic access to the aforementioned indicators, enabling third-party application development.
Finally, regarding the fourth guideline, our service provides the values of all indicators per topic (i.e., subject area).

Furthermore, many other principles that can facilitate the fair assessment of researchers are considered by \thiswork. First of all, it offers a wide range of indicators hence
its compliance towards fair principles is twofold: it captures as many perspectives of each researcher's performance as possible, but it also acts as a countermeasure to reduce the effects of attacks against individual indicators (since, according to Goodhart's or Campbell's law any popular indicator is bound to be abused). Even for scientific impact itself, \thiswork~offers a variety of indicators, each capturing impact from a different perspective (e.g., the \emph{popularity} indicator reflects the current hype of an article, while \emph{influence} reflects its overall impact~\cite{kanellos2019impact}). In addition, \thiswork~acknowledges not only the effort given by researchers to produce publications, but also other types of research works, while it also considers their efforts in practicing Open Science.

Finally, a major contribution of \thiswork~is that it offers researchers the opportunity to add useful information to their profiles, which only themselves could provide. For instance, they can define inactive periods in their careers (e.g., parental leaves, public service), something that allows for a fairer calculation of their academic age. 
But, more importantly, they are able to indicate their role in each of the research works in their track record (taking advantage of the CRediT taxonomy of roles). In an era when research activities often require diverse tasks to be completed, researchers undertake multiple responsibilities while, some of them, become more specialised in particular roles. Acknowledging the experience of a researcher in different roles is really important for a fair performance assessment, but this is something that has only recently become possible due to the creation of CRediT or other similar taxonomies (e.g., SCoRO\footnote{SCoRO: \url{http://www.sparontologies.net/ontologies/scoro}}). 

\section{Implementation}

In this section, we provide details about the main data sources we use to build and update the service (Section~\ref{sec:data}) and describe the system's key components and the technologies used (Section~\ref{sec:overview}).

\subsection{Data}
\label{sec:data}

Since many researcher-level indicators rely on citation network analysis, we constructed an interdisciplinary citation network, on which we calculate the respective 
indicator values.
To this end, we gather citation data and research work metadata from multiple data sources. In particular, the current dataset incorporates citations and research work metadata from the last snapshot (Dec 2021) of the Microsoft Academic Graph (MAG)~\cite{Sinha2015,Wang2020} and from recent snapshots of Crossref~\cite{Hendricks2020} (Dec 2021) and the OpenAIRE Research Graph~\cite{manghi2012} (Feb 2022), while we also gather citation data from the latest version (Jan 2022) of OpenCitations' COCI dataset.   

The aggregation of the previous datasets is a citation network, enriched with research work metadata on the nodes. It currently contains data for more than $140$M research works (publications and datasets) and $1.52$B distinct DOI-to-DOI relationships. It should be noted that we handle different DOIs as distinct research objects, therefore it is possible that multiple DOI entries may refer to the same object. Furthermore, since the publication year is required to compute some of the indicators, we discard DOIs for which this piece of information is missing in all our data sources.
In addition, we exclude entries with publication years greater than one year after the present
one, considering them as erroneous. Lastly, for all time-based analyses, we also set the value of the current year to the year following the year of the dataset's production. The full details about the data used for each indicator can also be  found in the ``Indicators'' page of \thiswork\footnote{\url{https://bip.imis.athena-innovation.gr/site/indicators}}, to aid users better understand their proper uses and limitations.

\subsection{System overview}
\label{sec:overview}

The data processing workflow of \thiswork~can be divided into the following components: 
\begin{itemize}
    \item \emph{Data aggregator.} It creates the citation network required for the calculation of most implemented researcher-level indicators by cleaning and integrating data from \thiswork's sources (see Section~\ref{sec:data}).  
    \item \emph{Work-level indicators calculator.} It is responsible for the calculation of the work-level indicators, which are used for the calculation of some of the researcher-level indicators (through aggregation). Number of works, citation counts, popularity, and impulse are calculated based on our own codes\footnote{\url{https://github.com/athenarc/yii2-scholar-indicators}}. For influence we have used an open-source Apache Spark-based implementation\footnote{BIP-Ranker library: \url{https://github.com/athenarc/Bip-Ranker}}.   
    \item \emph{Researcher-level indicators calculator.} This component calculates all researcher-level indicators. Some are calculated by aggregating the respective work-level indicators (e.g., citations, popularity, influence) while others are calculated from scratch (e.g., h-index, academic age). 
\end{itemize}

The prepared data are used by our Web-based user interface
(which is implemented in PHP using the Yii2
framework) to construct the researcher profile page and other relevant pages (e.g., the ``Indicators" page). 

\section{Demonstration}
\label{sec:demo}

During the demonstration, the audience will have the opportunity to become familiar with the basic fair research assessment initiatives and practices and to interact with \thiswork's user interface, exploring its functionalities. Although, the members of the audience will be free to interact with \thiswork~in any way they want, we have prepared a basic scenario to guide their interactions with the service in a way that showcases its different capabilities, in case they need guidance. The same scenario is designed to also guarantee that the work of the referees during the peer review phase will be uninterrupted and compliant with the mandates of the double-blind process. 

In particular, we have created a public profile page for a widely known researcher from the field of computer science. During the demonstration process, the members of the audience (following the guidelines of the presenter) are going to inspect the elements of this profile page, 
examining the values of different indicators and learning about their interpretation. Furthermore, they are going to interact with the facets to better understand different aspects of the work of the particular researcher. 

\section{Conclusion}

In this work, we have presented \thiswork, a Web-based service that offers researchers the opportunity to set up profiles that summarise their research careers taking into consideration well-established guidelines for fair research assessment, and facilitating the work of evaluators who want to be more compliant with the respective practices. 
As future work, we plan to add the support for additional types of research works and activities (e.g., software, peer-reviews) in \thiswork's profiles and to extend \thiswork~to offer profiles for research institutions (based on ROR\footnote{Research Organization Registry (ROR): \url{https://ror.org}} IDs) and activities (based on RAiDs\footnote{Research Activity Identifier (RAiD): \url{https://www.raid.org.au}}). 

\begin{acks}
This project has received funding from the European Union’s Horizon 2020 research and innovation programme under grant agreement No 101017452.

\begin{figure}[!h]
     \centering
         \includegraphics[width=0.1\linewidth]{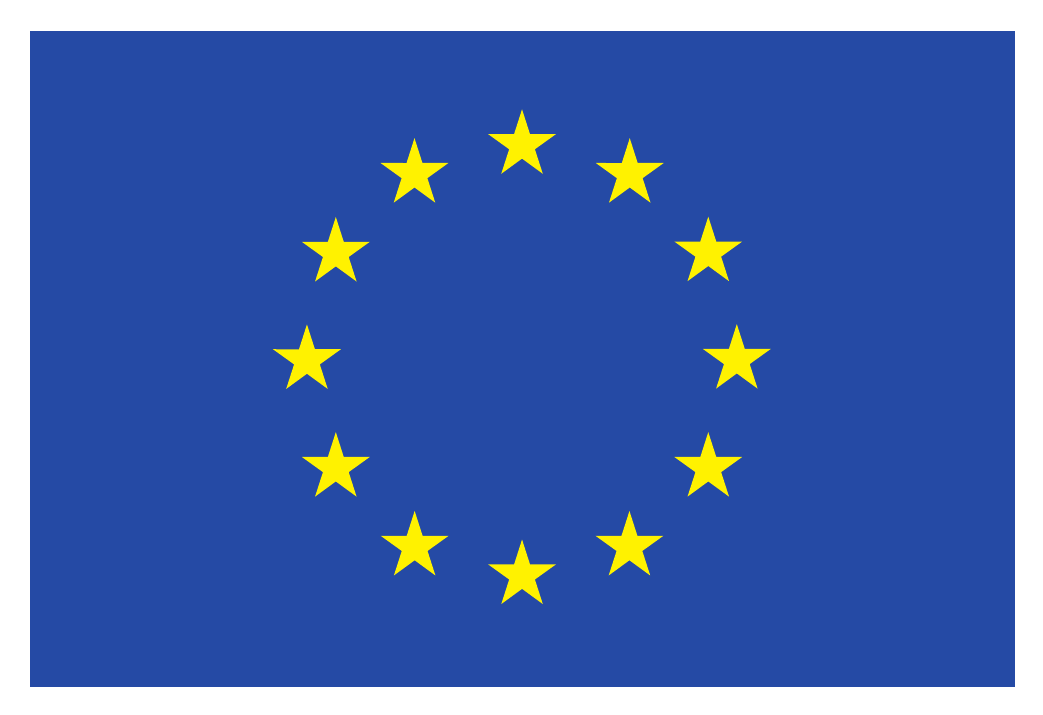}
\end{figure}

\end{acks}

\bibliographystyle{ACM-Reference-Format}
\bibliography{main}

\end{document}